\documentclass{emulateapj}
\usepackage{amssymb}

\begin{document}
\newcommand{\lya}{Lyman~$\alpha$}
\newcommand{\lyb}{Lyman~$\beta$}
\newcommand{\degpoint}{\mbox{$^\circ\mskip-7.0mu.\,$}}
\newcommand{\minpoint}{\mbox{$'\mskip-4.7mu.\mskip0.8mu$}}
\newcommand{\secpoint}{\mbox{$''\mskip-7.6mu.\,$}}
\newcommand{\sqdeg}{\mbox{${\rm deg}^2$}}
\newcommand{\squig}{\sim\!\!}
\newcommand{\subsun}{\mbox{$_{\twelvesy\odot}$}}
\newcommand{\et}{{\it et al.}~}
\newcommand{\Rs}{{\cal R}}

\def\ltsima{$\; \buildrel < \over \sim \;$}
\def\simlt{\lower.5ex\hbox{\ltsima}}
\def\gtsima{$\; \buildrel > \over \sim \;$}
\def\simgt{\lower.5ex\hbox{\gtsima}}
\def\propsima{$\; \buildrel \propto \over \sim \;$}
\def\simprop{\lower.5ex\hbox{\propsima}}
\def\arcs{$''~$}
\def\arcm{$'~$}

\title{POSSIBLE DETECTION OF LYMAN-$\alpha$ FLUORESCENCE FROM A DAMPED LYMAN-ALPHA SYSTEM AT REDSHIFT $Z\sim 2.8$\altaffilmark{1}}

\author{\sc Kurt L. Adelberger\altaffilmark{2}}
\affil{Carnegie Observatories, 813 Santa Barbara St., Pasadena, CA, 91101}

\author{\sc Charles C. Steidel}
\affil{Palomar Observatory, Caltech 105--24, Pasadena, CA 91125}

\author{\sc Juna A. Kollmeier}
\affil{Department of Astronomy, Ohio State University, McPherson Laboratory,
140 West 18th Avenue, Columbus, OH 43210}

\author{\sc Naveen A. Reddy}
\affil{Palomar Observatory, Caltech 105--24, Pasadena, CA 91125}

\altaffiltext{1}{Based on data obtained at the W.M. Keck
Observatory, which is operated as a scientific partnership between
the California Institute of Technology, the University of California,
and NASA, and was made possible by the generous financial support
of the W.M. Keck Foundation.}
\altaffiltext{2}{Carnegie Fellow}

\begin{abstract}
We have detected Lyman-$\alpha$ emission from a damped Lyman-alpha
system (DLA) that lies near the bright quasar HS1549+1919.
The DLA has the same redshift as HS1549+1919 and was discovered
in the spectrum of a faint QSO that lies $49''$ away
($380$ proper kpc).
The emission
line's luminosity, double-peaked profile, and small spatial separation 
from the DLA suggest that it may be fluorescent Lyman-$\alpha$ emission
from gas that is absorbing the nearby QSO's radiation.
If this is the case, our observations show that the DLA has
a size of at least $1\secpoint 5$ and that the QSO's luminosity
one million years ago was similar to its luminosity today.
A survey for similar systems within $\sim 1'$ of bright QSOs
would put interesting limits on the mean quasar lifetime.
\end{abstract}
\keywords{intergalactic medium --- quasars: absorption lines}

\submitted{Received 2005 June 10; Accepted 2005 September 2}
\shorttitle{FLUORESCENCE FROM A DLA}
\shortauthors{K.L. Adelberger et al.}

\section{INTRODUCTION \& DATA}
\label{sec:intro}
Although best known for
absorbing Lyman-$\alpha$ photons, intergalactic hydrogen also emits
them, at the rate of about one Lyman-$\alpha$ photon for every two
Lyman-continuum 
photons absorbed.
Average intergalactic clouds absorb photons so slowly that their 
emission is almost undetectable
(e.g., Hogan \& Weymann 1987; Gould \& Weinberg 1996), 
but optically thick clouds near QSOs should fluoresce far more brightly.
While obtaining data for a survey of galaxies
at redshifts $1.8\simlt z\simlt 3.5$ along the line of sight
to the bright QSO HS1549+1919 ($z\simeq 2.84$),
we discovered what
appears to be fluorescent Lyman-$\alpha$ emission
from a DLA that lies near the QSO.  We believe that this
is the first detection of fluorescence from intergalactic gas.\footnote{The
unique characteristic of the detection is the large distance (380 proper kpc, 
see below) between the fluorescent gas and the apparent source of the
ionizing photons.
Numerous authors have detected Lyman-$\alpha$ fluorescence from
the gas that lies within the same halo as claimed ionizing source
(e.g., Moller, Warren, \& Fynbo 1998; Bergeron et al. 1999; 
Fynbo, Thomsen, \& Moller 2000; Bunker et al. 2003; Moller, Fynbo, \& Fall 2004;
Weidinger et al. 2005).}

\begin{figure*}
\plotone{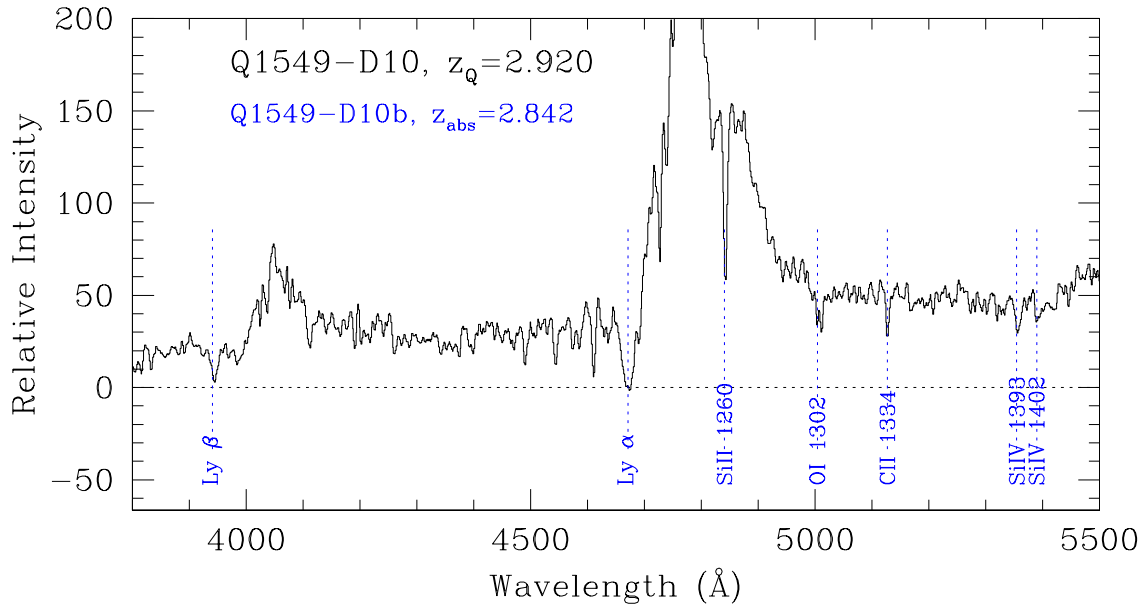}
\caption{\label{fig:d10spec} 
Spectrum of the QSO Q1549-D10 ($z=2.920$).  Absorption lines associated with
the DLA at $z=2.842$ are labeled.
}
\end{figure*}

The damped Lyman-$\alpha$ system has redshift $z=2.8420$
and column density $N_{\rm HI}=2.5\pm 0.5\times 10^{20}$ cm$^{-2}$.  
We identified it in the 
spectrum of the Lyman-break object Q1549-D10 (magnitude
$G_{\rm AB}=23.7$; coordinates $\alpha_{2000}=15^h51^m49^s.5$, 
$\delta_{2000}=19^o10'41''$;
redshift $z=2.920$), 
which is a faint QSO that
lies just $49''$ from the much brighter QSO HS1549+1919 ($G_{\rm AB}\sim 16$).
The spectrum, shown in Figure~\ref{fig:d10spec},
was obtained with 6 hours of integration time with the LRIS-B
spectrograph (Steidel et al. 2004) in multislit mode
on the Keck I telescope on 5 May 2005.
Its resolution is $\sim 3$\AA\ for point sources and $\sim 5$\AA\ for
extended sources.  The slit width was $1\secpoint 2$.
The redshift of the DLA appears to be the same as the redshift
of HS1549+1919, $z_Q\sim 2.84$, although the QSO's redshift has been
estimated from its broad emission lines and is therefore imprecise.\footnote{
In June 2005, we used NIRSPEC on the Keck II telescope
to measure a redshift of $z=2.8443\pm 0.0005$ for
the QSO's MgII emission line.  Richards et al. (2002)
report that the mean difference between the velocity
of MgII emission and a QSO's true recession velocity
is 97 km s$^{-1}$ and that the random scatter in the difference
is 270 km s$^{-1}$.  Neglecting the clustering of DLAs
around QSOs, we consequently estimate a $1\sigma$ confidence
interval for the difference between the QSO and DLA velocities
of $\delta v = 80\pm 270$ km s$^{-1}$.}

\begin{figure}
\plotone{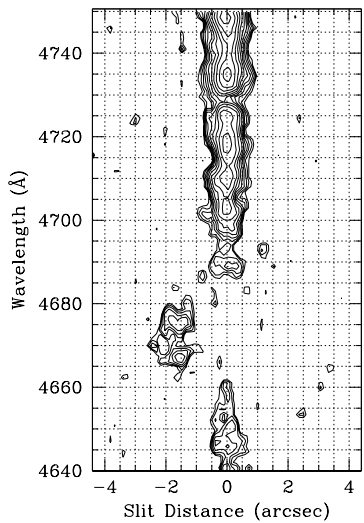}
\caption{\label{fig:spec_contour} 
Two dimensional spectrum of D10.  Wavelength increases from the bottom
to the top of the figure, and spatial position varies from left to right.
The damped Lyman-$\alpha$ absorption line in the QSO spectrum at
$x=0$, $\lambda\sim 4670$ is accompanied by an emission line with a small
spatial offset.
}
\end{figure}

\begin{figure}
\plotone{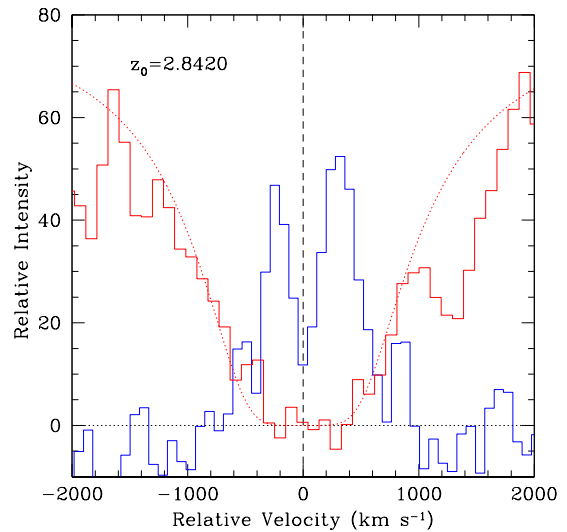}
\caption{\label{fig:velplot} 
Velocity profiles of the double-peaked emission line and of the
nearby damped absorption.  Solid lines show our data; the dotted line
shows our fit to the absorption profile.
All velocities are relative to
the DLA metal lines visible in Figure~\ref{fig:d10spec}.
}
\end{figure}

The spectroscopic slit centered on Q1549-D10 had a length 
of $15''$.  The two-dimensional spectrum shows a
broad, double-peaked emission line roughly $1\secpoint 5$ 
to the east of Q1549-D10 (Figure~\ref{fig:spec_contour}).  
Comparing the line's spatial width ($0\secpoint 84$ FWHM) to the width of the nearby
point source Q1549-D10 ($0\secpoint 72$ FWHM), we estimate an intrinsic size for
the emitting region of roughly one-half arcsecond.\footnote{
By repeatedly adding random noise to each pixel's observed flux 
and recalculating
the widths of Q1549-D10 and the emitting region, we estimate
a 90\% confidence interval of $0\secpoint 3<\Delta\theta<0\secpoint 7$
for the emitting region's intrinsic (i.e., quadrature subtracted) FWHM.}
We interpret the emission as Lyman-$\alpha$ from the DLA,
since its central wavelength is precisely
what one would expect for Lyman-$\alpha$ at the redshift of
the DLA's metal lines (Figure~\ref{fig:velplot}).


\begin{figure}
\plotone{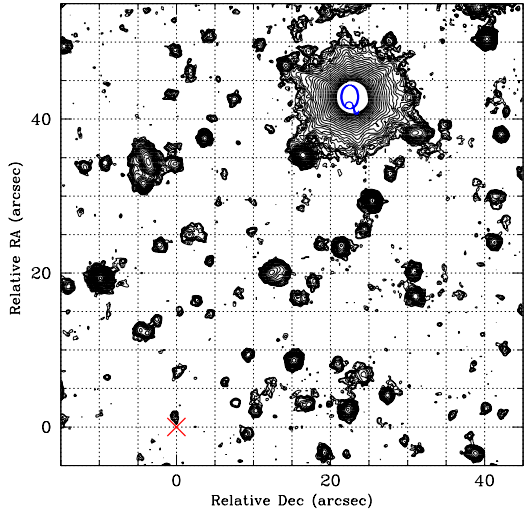}
\caption{\label{fig:G_contour} 
$G$-band image of the field.  Q1549-D10, at position $(0,0)$,
has been subtracted.  The Lyman-$\alpha$ emitting region lies
just above it on this plot.  The bright QSO HS1549+1919 is
labeled ``Q''.
}
\end{figure}

The relative spatial positions of
HS1549+1919, Q1549-D10, and the Lyman-$\alpha$ emission region are shown
in Figure~\ref{fig:G_contour}.  This is a section of a 2960-second $G$-band  
image that we obtained with LRIS-B on 6 April 2005.  The $U_n$ and ${\cal R}$
images discussed below were obtained on the same run,
with exposure times of 3600s and 4800s respectively.
Seeing was $0\secpoint 9$ in each band.

\section{INTERPRETATION}
\subsection{Theoretical expectations for fluorescence}
The proximity of the DLA to HS1549+1919 suggests that
the emission could be fluorescent, fueled by the absorption
of the bright QSO's Lyman-continuum photons by the DLA's
optically thick gas.
If the Lyman-$\alpha$ emission were powered by the QSO, we would expect it to
have the following three properties.  

(1) The Lyman-$\alpha$ line should have a width of roughly~8
times the velocity dispersion of the absorbing gas.  Since Lyman-$\alpha$
photons can escape an optically thick 
cloud only when they are scattered away from
the resonant frequency,
the Lyman-$\alpha$ photons that emerge will come from 
the $\sim\pm 4\sigma$ wings
of the Doppler profile (e.g., Zanstra 1949; Osterbrock 1962;
Urbaniak \& Wolfe 1981; Gould \& Weinberg 1996; Zheng \& Miralda-Escud\'e 2002b;
cf. Cantalupo et al. 2005).
The DLA's marginally resolved CII $\lambda$1334 absorption line implies
a velocity width of $\sigma\sim 60$ km s$^{-1}$, so
the emission should be spread over a range of velocities
$\Delta v\sim 500$ km s$^{-1}$ if the absorbing gas
has a velocity dispersion similar to the DLA's.  
The emission would emerge in two peaks
with this separation if the gas were homogeneous and would
be more continuously distributed throughout $\Delta v$ otherwise.
(The expected velocity dispersion would be significantly reduced, however,
if all the ionizing photons were absorbed by a single cold clump
orbiting in the DLA potential;
see
Gould \& Weinberg 1996, Zheng \& Miralda-Escud\'e 2002b, Cantalupo et al. 2005,
and Kollmeier et al. 2005, in preparation, for further discussion.)  

(2) The Lyman-$\alpha$ emission should emerge
from the side of the DLA that lies closest to the QSO.
According to Gould \& Weinberg (1996; \S~3), 
more than $98$\% of the Lyman-$\alpha$
photons will escape from the side the Lyman-continuum photons entered
for column densities $N_{\rm HI}>10^{20}$ cm$^{-2}$.
Since the DLA will not begin
to fluoresce brightly until the
ionization front from the QSO has penetrated deeply into it,
most of the way to the center, 
the fluorescent emission should emerge from
near the center of the DLA.  Shielded neutral gas should
extend from the site of emission in the direction
opposite the QSO. Figure~\ref{fig:schematic} illustrates the situation.
See the appendix for further discussion.

\begin{figure}
\plotone{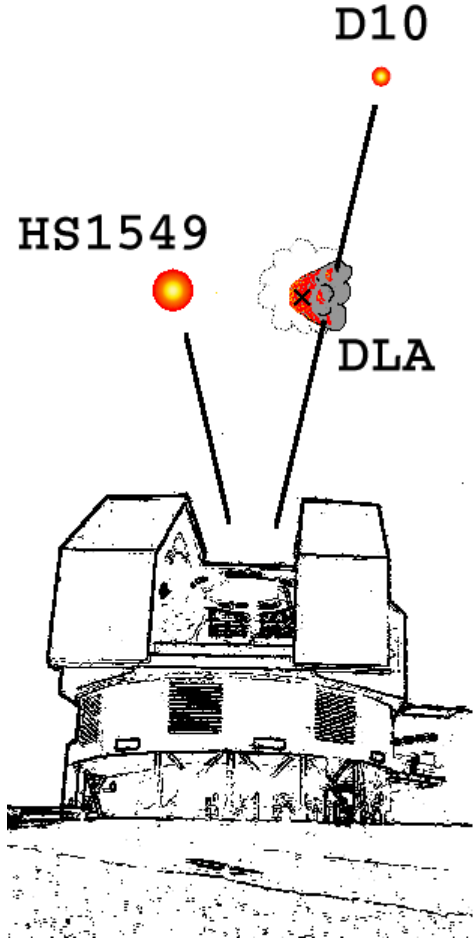}
\caption{\label{fig:schematic} 
Cartoon view of the proposed geometry.  
The DLA's original size (dotted lines) has been reduced
to the grey region
by intense radiation from the nearby QSO HS1549+1919.
Lyman-$\alpha$ fluorescence emerges from the side of the DLA
that faces the QSO.
This surface lies close to the stars at the original center of the DLA
(black cross).
Light from the background QSO Q1549-D10 passes through gas that
is shielded from the QSO, and we observe damped absorption.
}
\end{figure}

(3) Its surface brightness should be 
\begin{equation}
\Phi = (1+z)^{-3}\eta_{\rm thick} F_c / \pi
\label{eq:phi}
\end{equation}
where $\eta_{\rm thick}\simeq 0.6$ (e.g., Gould \& Weinberg 1996)
is the ratio of Lyman-$\alpha$ photons emitted to Lyman-continuum
photons absorbed and
$F_c$ 
is the flux (cm$^{-2}$ s$^{-1}$) of hydrogen-ionizing
photons into 
the DLA.  
The following calculation implies that the expected fluorescent
surface brightness is
$\Phi \sim 2.7\times 10^{-5}$ photons cm$^{-2}$ s$^{-1}$ arcsec$^{-2}$,
which implies a luminosity of
$L \simeq 1.1\times 10^{-16}$ erg cm$^{-2}$ s$^{-1}$ if the fluorescing
region has a size of 1 arcsec$^{2}$.
Before the QSO's radiation hits, the ionizing flux is
\begin{equation}
F_c^0=\pi\int_{\nu_0}^\infty d\nu\,\frac{J_{\rm bg}(\nu)}{h\nu}
\label{eq:fc0}
\end{equation}
where $J_{\rm bg}$ is the ambient radiation field.
The value of $F_c^0$ is $\sim 1.6\times 10^5$
for plausible values of $J_{\rm bg}$, e.g., 
$J_{\rm bg}\sim 5\times 10^{-22}(h\nu/13.6{\rm eV})^{-1.8}$
erg s$^{-1}$ cm$^{-2}$ sr$^{-1}$, which we will assume throughout. 
After the radiation hits, the flux will increase by the factor
\begin{equation}
g = 1+\frac{10^{-0.4(48.60+m_{912})}}{(1+z)\pi J_{\rm bg}(c/912{\rm\AA})}\Bigl[\frac{d_L(z)}{r_Q}\Bigr]^2 \cos\theta,
\label{eq:defG}
\end{equation}
where $m_{912}$ is the QSO's AB magnitude at rest-frame 912\AA,
$d_L(z)$ is the luminosity distance to redshift $z$, $r_Q$ is the
proper distance between the DLA and QSO, $\theta<\pi/2$ is the angle between
the vector to the QSO and the vector normal to the DLA's surface,
and $J_{\rm bg}$ has units
erg s$^{-1}$ cm$^{-2}$ Hz$^{-1}$.
Equation~\ref{eq:defG} assumes that the spectra of the QSO and of the
ambient radiation have the same shape shortwards of 912\AA.
Scaling a HIRES spectrum of HS1549+1919 
(W. Sargent 2005, private communication) to match the QSO's observed
$U_n$ magnitude, $U_n=16.9$, we estimate a continuum AB magnitude
at 912\AA\ of $m_{912}\simeq 16.7$.
This implies
$g\simeq 4850 \cos\theta $ if the QSO and DLA have separation $49''$
and identical redshifts.  
For a spherical DLA with half-moon illumination, 
the effective value of $g$ is $\langle g \rangle = 4 g_{\rm max} (3\pi)^{-1}$,
where $g_{\rm max}$ is the value of $g$ for $\theta=0$.
We consequently adopt $g\sim 2000$.
The value of $\Phi$ quoted at the
beginning of the paragraph follows from inserting
$F_c = gF_c^0$ into equation~\ref{eq:phi}.

\subsection{Comparison to observations}
The observed characteristics of the Lyman-$\alpha$ line agree surprisingly
well with these expectations.  The line is broad,
FWHM$\sim 990$ km s$^{-1}$, and appears to consist of two peaks
with separation
$\Delta v=530$ km s$^{-1}$.
Its luminosity of $L=2.1\pm 0.5\times 10^{-17}$ erg s$^{-1}$ cm$^{-2}$
and spatial extent of roughly one-half arcsecond imply a 
surface brightness similar to the expectation
$\Phi=1.1\times 10^{-16}$
erg s$^{-1}$ cm$^{-2}$ arcsec$^{-2}$.
The emission is produced
by hydrogen that lies between the neutral gas in the spectrum
of Q1549-D10 and the bright illuminating source HS1549+1919, and
its offset from the neutral gas, $\sim 1\secpoint 5$, is consistent with
theoretical expectations for the size of DLAs
(e.g., McDonald \& Miralda-Escud\'e 1999; Haehnelt, Steinmetz, \& Rauch 2000;
Zheng \& Miralda-Escud\'e 2002a).

Although suggestive, the good agreement with these expectations does not prove
that the Lyman-$\alpha$ emission is powered by the distant QSO.
Similar emission-line characteristics could arise by chance
if the line were powered by a local source.  This possibility
should be taken seriously, as we now show.

\section{DISCUSSION}
Our detection of the emitting region in the $G$ and ${\cal R}$ bands
(Figure~\ref{fig:G_contour}; Table~\ref{tab:dla}) implies that
it has substantial luminosity in the continuum.  Comparing its 
observed $G$-band magnitude, $G_{\rm AB}=26.8\pm 0.2$, to the
flux detected in the emission line, we estimate that the observed-frame
equivalent width of Lyman-$\alpha$ is $W_\lambda\sim 275\pm 75$\AA. 
This is several times smaller than would be expected if
the emission were powered solely by the distant QSO
(see, e.g., the calculation in the appendix).
We are forced to conclude that the continuum receives a substantial
contribution from some other source, presumably the stars that are associated
with the DLA. 

\begin{deluxetable}{cccccc
}\tablewidth{0pc}
\scriptsize
\tablecaption{DLA characteristics}
\tablehead{
        \colhead{$\alpha(2000)$\tablenotemark{a}} &
        \colhead{$\delta(2000)$\tablenotemark{a}} &
        \colhead{$N_{\rm HI}$\tablenotemark{b}} &
        \colhead{$W_\lambda$\tablenotemark{c}} &
        \colhead{$G$\tablenotemark{d}} &
        \colhead{${\cal R}$\tablenotemark{d}}
}
\startdata
15 51 49.6 & 19 10 41  &  $2.5\pm 0.5$  &  $275\pm 75$ & $26.8\pm 0.2$ & $26.5\pm 0.2$ \\
\enddata
\tablenotetext{a}{Coordinates of Lyman-$\alpha$ emitting region.}
\tablenotetext{b}{Column density in the spectrum of Q1549-D10, $10^{20}$ cm$^{-2}$}
\tablenotetext{c}{Observed-frame equivalent width of Ly-$\alpha$ emission, \AA}
\tablenotetext{d}{AB magnitude of the Lyman-$\alpha$ emitting region.}
\label{tab:dla}
\end{deluxetable}

This does not contradict the idea that the Lyman-$\alpha$ line is
powered primarily by fluorescence (cf. Fynbo, Thomsen, \& M\/oller 2000).  
The DLA's strong metal lines
imply that it ought to contain stars, and the calculation in the appendix
suggests that the fluorescence should be produced by material that
lies only $\sim 0\secpoint 1$ from the center of the DLA potential, which
is presumably where any stars would be found.
One could have anticipated its relatively small equivalent width.

Could local emission from the DLA power the entire Lyman-$\alpha$ line?
Although we cannot rule out
the possibility, it seems implausible to us.
First, this type of Lyman-$\alpha$ line is rarely produced by stars.
Only 2\% of randomly selected Lyman-break galaxies at this redshift have
a Lyman-$\alpha$ emission line with so large an equivalent width
(Steidel et al. 2000).
Velocity widths ${\rm FWHM}\sim 1000$ km s$^{-1}$ are observed 
only in the merging or active galaxies of the Lyman-break sample, 
and we see no sign of either
the rapid star formation that would accompany the former or
the other broad emission lines (CIV, etc.) that would accompany the latter.
Although broad and strong Lyman-$\alpha$ lines are (not surprisingly)
more common among 
galaxies discovered in Lyman-$\alpha$ surveys (e.g., Francis et al. 2004),
fluorescence remains the best 
explanation for the observed
double-peaked profile.
Second, regardless of its stellar content, the DLA is optically thick
to Lyman-continuum photons and lies near a very bright source.
It is guaranteed to fluoresce unless the QSO's radiation is beamed
or has not yet had time to reach it.

Isotropically emitted Lyman-continuum photons from the QSO will have reached the DLA at the time
of observation as long as the QSO was shining at the earlier time
$\Delta t = [(R^2+Z^2)^{1/2}+Z]/c$, where $R\equiv D_A(\bar z)\theta$ 
and $Z\equiv [D_C(z_{\rm DLA})-D_C(z_Q)]/(1+\bar z)$ are the proper
separations of the QSO and DLA perpendicular and parallel to
the line of sight, $D_A$ is the angular diameter distance,
$D_C(z)$ is the comoving distance to redshift $z$, and 
$\bar z\equiv (z_{\rm DLA}+z_Q)/2$ (e.g., Adelberger 2004).
If the DLA and QSO have exactly the same redshift, so that
$Z=0$, then photons will reach the DLA roughly $1.2$ Myr 
after they were emitted.  Since the DLA will probably not begin
to fluoresce brightly until $\sim 0.1$ Myr after it is first illuminated
(see the appendix),
we would expect to see fluorescent emission as long as the
QSO has been radiating for at least $1.3$ Myr. 
The minimum lifetime could be somewhat longer or shorter, depending on
the actual value of $Z$, but if $Z$ were significantly different from 0
the luminosity of the Lyman-$\alpha$ line would be too large to be explained
by the QSO's radiation.\footnote{The nominal $1\sigma$ uncertainty in
the velocity difference between the QSO and DLA, 270 km s$^{-1}$
(Richards et al. 2002), 
corresponds to a proper distance of $920$ kpc.  If the DLA lay this
far behind the QSO, its expected surface brightness would decrease
by a factor of $\sim 7$ but its illuminated size would nearly double.
The predicted Lyman-$\alpha$ flux would be consistent with our observations as
long as the DLA were somewhat larger than we have assumed.
In this case the implied length of the QSO lifetime would be 
$\simgt 6$ Myr.
If the DLA lay in front of the QSO by 920 kpc, however, only a small fraction
of its illuminated surface would be visible to us and the flux would
be far lower than we have observed.}

Regardless of the origin of the line, we can conclude that
this DLA has a size of at least $\sim 1\secpoint 5$.  If $1\secpoint 5$ is the
typical radius of DLAs, then their observed number per unit redshift
interval, $dN/dz\sim 0.2$ (e.g., Boissier, P\'eroux, \& Pettini 2003),  
corresponds to a comoving number density of $\sim 10^{-1} h^3$ Mpc$^{-3}$.
The average QSO would therefore have $\sim 1$ DLA within $1.3h^{-1}$ Mpc
(i.e., $1'$ projected) if DLAs and QSOs were spatially uncorrelated
and $\sim 18$ within the same radius if the QSO-DLA correlation function
were similar to the QSO-LBG correlation function, 
$\xi(r)\sim (r/5h^{-1}{\rm Mpc})^{-1.6}$ (Adelberger \& Steidel 2005).\footnote{
Including optically thick clouds with lower (i.e., sub-DLA) column densities
would increase the numbers significantly, although these clouds might not
survive for long near the QSO.}
Fluorescing systems similar to the one we have observed should therefore
be common if QSOs live longer than a few Myr and optically thick clouds
are not destroyed by their radiation.  Systematic surveys for them
(e.g., M\/oller, Warren, \& Fynbo 1998; Fynbo et al. 2000;
Francis \& Bland-Hawthorn 2004)
should provide interesting constraints on the lifetimes of QSOs and
the nature of 
intergalactic clouds.

\bigskip
\bigskip
KLA was educated by his discussions with
T. Abel, A. Gould, M. Rauch, D. Weinberg,
and (especially) J. Miralda-Escud\'e, who suggested a 
power-law density profile for intergalactic clouds and
pointed out that it implied equation~\ref{eq:ri_o_rf}.
R. Simcoe tracked down a raw HIRES spectrum and
confirmed the redshift of HS1549+1919.
We are grateful to 
our collaborators D. Erb, M. Pettini, and
A. Shapley for their enormous contributions to our ongoing survey.

\appendix

\section{RESPONSE OF A DLA TO A BLAST OF RADIATION}
In the text we assumed that the ionization front from the
QSO would advance most of the way to the center of the DLA,
that the DLA would not glow brightly in Lyman-$\alpha$ until
the ionization front was close to its final radius, and
that the rest-frame equivalent width of fluorescent Lyman-$\alpha$
would be at least $200$--300\AA\ unless the continuum received
a contribution from some other source.  The last two assumptions should
be true in general.  The first will be true if the gas is smooth
and centrally concentrated but may be false otherwise. 
Our reasoning is illustrated below with a simple physical model.
Please recall that observed damped Lyman-$\alpha$ systems show
a wide range of kinematic properties, from the simple to the 
clumpy and complex (e.g., Prochaska \& Wolfe 1997; Prochaska \& Wolfe 1998).
The model's predictions should not be taken as anything 
other than a very rough guide.

We begin with a brief overview of the physical situation.
Before it is struck by radiation from the QSO,
the hydrogen at the optically thick edge of the DLA
will presumably be in rough thermal and ionization equilibrium
with the ambient radiation field, recombining as fast as it is
ionized and emitting energy as fast as energy is absorbed.
If the QSO has the same spectrum as the ambient background,
its radiation will increase both the heating and ionization
rates by the same factor $g\sim 2000$, driving the cloud wildly
out of equilibrium.\footnote{The factor $g\sim 2000$, derived in the main body
of the text, is appropriate for a DLA that lies $0.38$ proper Mpc from
a QSO with the magnitude of HS1549+1919.}  
An ionization front will advance into the cloud.  At first
the recently ionized material will not be able to cool from
its initial temperature of $\sim 50000$K (Abel \& Haehnelt 1999).
The front will advance until it encounters material dense enough
to recombine as fast as it is ionized.  It will stall.  The
cooling rate will now exceed the heating rate, and the ionized material
will cool until it reaches the same temperature it had before
the radiation blast arrived.\footnote{Since the heating and ionization
rates depend on density in the same way and on temperature in
different ways, and since (according to our assumptions)
each is increased by the same factor $g$
when the QSO's radiation arrives, the initial and final equilibrium
temperatures must be the same.}
The cloud will finally arrive at the equilibrium situation envisioned by
Gould \& Weinberg (1996).

\subsection{Stalling Radius}
We estimate the stalling radius for the ionization front by
assuming that the cloud is centrally concentrated, with
a mean density profile that varies with radius as $r^{-\gamma}$.
In this case the integrated 
recombination rate for all the material
exterior to radius $r$ along a column of fixed cross-section
is proportional to $r^{1-2\gamma}$.
An increase in the ionization rate by a factor of $g$ will
be balanced by recombinations only when the radius has shrunk by the factor 
\begin{equation}
\frac{r_{\rm initial}}{r_{\rm final}} = g^{1/(2\gamma-1)},
\label{eq:ri_o_rf}
\end{equation}
or $\sim 13$ if $g\sim 2000$ and the cloud has an isothermal profile
(i.e., $\gamma=2$).  
The QSO radiation that is pointed directly at the center of the DLA
will therefore advance most of the way to the center before it stalls.
If the initial ionized edge of the DLA lies at a radius of
$1\secpoint 5$, the stalling radius will be $\sim 0\secpoint 1$.
Since the recombination rate (and hence Lyman-$\alpha$ luminosity)
is proportional to $g(r/r_{\rm final})^{-(2\gamma-1)}$ when the
ionization front is at radius $r$, and $2\gamma-1\sim 3$,
the Lyman-$\alpha$ emission will not be bright enough for
easy detection until the ionization front is quite close to its stalling radius.
We conclude that the separation between the Lyman-$\alpha$ fluorescence and the
center of the DLA is unlikely to be resolvable from the ground.
Clumping in the gas will not alter this conclusion as
long as the strength of clumping $\langle\rho^2(r)\rangle/\langle\rho(r)\rangle^2$ 
does not depend strongly on $r$.

\subsection{Stalling time}
The same model provides a rough estimate of the time required for
the cloud to reach its new equilibrium.  Since the column density
of material exterior to radius $r$ is proportional to $r^{1-\gamma}$,
the QSO will have to ionize a neutral hydrogen column of
$N_{\rm HI}=N_i [(r_{\rm final}/r_{\rm initial})^{1-\gamma}-1] = N_i [g^{(\gamma-1)/(2\gamma-1)}-1]$ 
as it advances
from the initial to the final radius.  $N_i$, the total (neutral plus ionized)
hydrogen column density
exterior to $r_{\rm initial}$, is likely to be roughly $10^{20}$ cm$^{-2}$,
since the optically thick edge of the cloud has a neutral hydrogen column density
of $N_{\rm HI}\sim 10^{17}$ cm$^{-2}$ and the typical
neutral fraction at this radius is 
$\sim 10^{-3}$
(Zheng \& Miralda-Escud\'e 2002a).
Because the ionization rate
is $\dot N_{\rm HI} = g F_c^0$, where $F_c^0$ 
is the inward flux of hydrogen-ionizing photons (cm$^{-2}$ s$^{-1}$)
in the ambient radiation
field (equation~\ref{eq:fc0}),
the rough time scale for reaching the equilibrium radius is 
$t_{\rm grow} \sim N_{\rm HI}/\dot N_{\rm HI} \sim g^{-\gamma/(2\gamma-1)} N_i/F_c^0$
for $g\gg 1$.  The required time is $t_{\rm grow}\sim 1.2\times 10^5$ yr
for $g=2000$, $N_i=10^{20}$ cm$^{-2}$, and $F_c^0 = 1.6\times 10^5 {\rm cm}^{-2}\, {\rm s}^{-1}$.  
This is almost short enough
to ignore in the present case, but could be significant for DLAs that
are less intensely illuminated.  Uncertainties in $N_i$ mean that
this time-scale should not be ignored in general.
Clumping in the gas does not alter the expected value of
$t_{\rm grow}$ but causes its actual value vary randomly
among different rays towards the center of the DLA.
These variations could be large for DLAs with significant sub-structure.

\subsection{Temperature and Lyman-$\alpha$ equivalent width}
In ionization equilibrium, the photoionization rate at radius $r$
will equal the recombination rate at the same radius.
Most Lyman-continuum photons will therefore be absorbed
in regions with low neutral fraction ($\sim 0.01$--$0.1$), since these
are the regions with the highest recombination rates.  
The energy
of the Lyman-continuum photons will have to be dissipated by
mostly ionized gas.  
Assume momentarily that the gas is completely ionized.
In this case, recombination and Brehmsstrahlung
would be the main cooling processes at the relevant
range of densities and temperatures.
According to the rate coefficients collected in Tables~1 and~2
of Katz, Weinberg, \& Hernquist (1996), energy losses from
H and He recombination will exceed those from Brehmsstrahlung
by a factor of around $\sim 2$--$3$, with the exact factor depending
on the temperature and on the dominant ionization state of He.
These same coefficients imply that the time-scale $t_{\rm cool}\equiv E/\dot E$
for recombination cooling should be 2--3 times 
the recombination time-scale $t_{\rm rec}\equiv n_{\rm HII}/\dot n_{\rm HII}$.
Since the recombination time-scale is proportional to
$r^{\gamma} T^{0.7}$, and had initial value
$t_{\rm rec}\sim N_i/\zeta\sim 20$Myr, 
the cooling time at the stalling radius for completely ionized gas 
is 2--3 times
longer than $20 g^{-\gamma/(2\gamma-1)} (T_s/T_i)^{0.7}$ Myr,
or $t_{\rm cool}\sim 1$ Myr.  Here $T_i\sim 10000$K and $T_s\sim 50000$K
are the temperatures in the ionized skin of the DLA before and after
the QSO's radiation arrives.  Figure~1 of Katz et al. (1996) shows
that line cooling is roughly two orders of magnitude faster
than recombination cooling for predominantly neutral gas
at these temperatures,
and so the actual cooling time could be significantly shorter than
$1$ Myr if (as seems likely) the residual neutral fraction in the
absorbing gas is greater than $\sim 0.01$.
We conclude that the gas is likely to have reached
a nearly equilibrium temperature, but
uncertainties in $N_i$ mean that one should not
always assume that this is the case.

Because recombination cooling proceeds 2--3 times faster than
Brehmsstrahlung cooling, roughly 70\% of the energy absorbed from
Lyman-continuum photons would be carried out of the cloud by 
recombination photons if the absorbing
gas were completely ionized.  With
40\% of the escaping recombination energy in the form of
Lyman-$\alpha$ photons (Gould \& Weinberg 1996), 
the Lyman-$\alpha$ luminosity would
equal $\sim 30$\% of the absorbed Lyman-continuum luminosity
in this case.
Gould \& Weinberg (1996) show that the fraction rises
to $\sim 50$\% if the absorbing gas is predominantly neutral.
The actual fraction will be bracketed by these two values.
The remaining energy will emerge either in other H/He recombination lines
or in a continuum produced by two-photon decay and
by the first (i.e., free-bound) photons of the recombination
cascades.  Much of this remaining energy will emerge
in the wavelength range $1100\simlt\lambda/{\rm\AA}\simlt 2000$ (i.e., 6--11 eV)
covered by our $G$ and ${\cal R}$ broad-band images,
but even if all of it were concentrated in that range it would not
be enough relative to Lyman-$\alpha$ to explain the relatively small
observed equivalent width of the Lyman-$\alpha$ line.
This conclusion holds equally well for smooth and clumpy gas.

\end{document}